\documentclass[12pt]{article}

\usepackage{amstext,amsmath,amssymb,amsfonts,bbm}
\usepackage[latin1]{inputenc}
\usepackage{epsfig}
\usepackage{hyperref}
\usepackage{amsthm}
\usepackage{subfigure}
\usepackage{color}
\usepackage{multirow}
\newcommand{\bea}{\begin{eqnarray}}	
\newcommand{\eea}{\end{eqnarray}}
\newcommand{\be}{\begin{equation}}	
\newcommand{\ee}{\end{equation}}
\newcommand{\cG}{{\cal G}}
\newcommand{\cA}{{\cal A}}
\newcommand{\ra}{\rangle}
\newcommand{\la}{\langle}

\newtheorem{lemma}{Lemma}

\begin{document}

\title{A Diagrammatic Equation for\\ Oriented Planar Graphs}

\author{Razvan Gurau\footnote{Perimeter Institute for Theoretical Physics Waterloo, ON, N2L 2Y5, Canada}}

\maketitle

\begin{abstract}
\noindent In this paper we introduce a diagrammatic equation for the planar sector of square non hermitian random matrix models strongly reminiscent of Polchinski's equation in quantum field theory. Our fundamental equation is first obtained by a graph counting argument and subsequently derived independently by a precise saddle point analysis of the corresponding random matrix integral. We solve the equation perturbatively for a generic model and conclude by exhibiting two duality properties of the perturbative solution.
\end{abstract}

\section{Introduction}

Random matrix models have wide applications in subjects ranging from statistical physics
\cite{Guhr:1997ve} to two dimensional quantum gravity \cite{Di Francesco:1993nw}. They have long been related to triangulations of two dimensional surfaces \cite{Brezin:1977sv} and various combinatorial counting problems \cite{Bouttier:2002iw, Di Francesco:2002ru, ZinnJustin:2003kw} have been addressed with their help. Techniques inspired by quantum field theory have been adapted to such models \cite{Brezin:1992yc}, and non identically distributed random matrix models proved crucial in the study of non commutative quantum field theories \cite{GrWu1,GrWu2,DGMR,GGR}. 

For a large class of random matrix models, in the limit of large matrices (the so called ``large $N$ limit''), the behavior of the partition function (and all correlation functions) is dominated by a saddle point. The planar sector \cite{Brezin:1977sv} is the first contribution to the evaluation of the random matrix integral by the saddle point method. The study of this saddle point is best addressed by means of the resolvent function which obeys a certain quadratic equation. In the large $N$ limit the resolvent is the generating function of planar graphs with one vertex, thus having a very natural graphical interpretation.

Rectangular non hermitian random matrices been extensively studied (see \cite{Di Francesco:2002ru} and references therein). However the treatment presented \cite{Di Francesco:2002ru} is not immediately applicable to square matrices: it is singular already for a free model (as it will be explained in detail in section \ref{sec:saddle}). The solution implemented in \cite{Di Francesco:2002ru} is to study rectangular non hermitian matrices and take the limit of square matrices {\it after} having taken the large $N$ limit.

In this paper we propose a new approach to square non hermitian random matrices. Our approach does not suffer from the singularities of \cite{Di Francesco:2002ru} (and sheds light on there origin). A detailed analysis allows us to derive a tower of equations obeyed by the {\it coefficients} of resolvent of such models both by a graph counting argument (inspired by the Polchinski equation) and by a saddle point analysis.  We will solve these equations in perturbations and prove certain duality properties of the solution. Our results encode the census of planar diagrams with one vertex generated by an arbitrary non hermitian random matrix model (which can be translated into the census of planar connected amputated graphs with one external face). Non hermitian matrix models have been studied in \cite{Am1,Am2} by the method of the ``loop insertion equation'' (very similar to our equation for the resolvent) which can be derived either by a combinatorial argument (for the resolvent itself) or by a functional manipulation. In \cite{Am1,Am2} the equation is solved by contour integrations (allowing rapid access to large order behavior of the solution) rather than trough the ``brute force'' approach we take here. Their results provide a check of our method and suggest possible roads to generalize our tower of equations for non planar graphs.

In higher dimmensions, random matrix models generalize to group field theories \cite{laurentgft,iogft}, associated to cellular complexes \cite{Color} with boundary \cite{PolyColor}. A generalized notion of planarity adapted to group field theory has been proposed \cite{GFTplanar}, and various power counting results \cite{GFTplanar, Sefu1,Sefu2} have been established. Our long term goal would be to generalize the results of this paper for such theories.

This paper is organized as follows. In section \ref{sec:generic} we present the non hermitian random matrix models and the relation between the resolvent and the connected planar functions with one external face. In section \ref{sec:counting} we derive our fundamental equations by a graph counting argument, and recover them in section \ref{sec:saddle} by a saddle point analysis. Section \ref{sec:sol} provides the perturbative solution of these equations. In section \ref{sec:dualities} we exhibit two duality properties of the perturbative solution. Section \ref{sec:conc} draws the conclusions of our work.
 
\section{Non Hermitian Random Matrix Models}\label{sec:generic}

In this section we review in some detail the non hermitian random matrix models. We use this opportunity to introduce our notations, starting with the most important: troughout this paper we will denote 
\bea
C^{n}_{p}=\frac{n!}{p!(n-p)!}\; ,
\eea
the binomial coefficient.

Let $M$ be an square ($N\times N$) non hermitian ($M\neq M^{\dagger}$) matrix. A non hermitian random matrix model is a probability measure 
\bea \label{eq:definition}
&& d\mu^{}_{V(\alpha M^{\dagger}M)}=\Big{(}\frac{N}{2\pi\imath}\Big{)}^N\Big{[} \prod_{ab} dM^*_{ab} \; dM_{ab} \Big{]} \; e^{-N\Big{(} \text{Tr}(M^{\dagger}M) + \text{Tr}[V(\alpha M^{\dagger} M)]\Big{)}} \nonumber\\
&& V(\alpha M^{\dagger}M )= - \sum_{p=1}^{\infty} \frac{1}{p} \; g_p \; \alpha ^p \; (M^{\dagger}M)^p \; .
\eea
with $\alpha$ some small coupling constant\footnote{The exponent $\text{Tr}(M^{\dagger}M) + \text{Tr}[V(\alpha M^{\dagger} M)]$ in 
eq. (\ref{eq:definition}) is called the action of the model. A rescaling of both $M$ and $M^{\dagger}$ by $\sqrt{\alpha}$ brings it in the more familiar form $ \frac{1}{\alpha}\text{Tr}(M^{\dagger}M) + \text{Tr}[V(M^{\dagger} M)]$.}. 
The normalization is chosen such that at $\alpha=0$, $d\mu_0$ is a normalized gaussian measure. The partition function and the 
correlations of this measure are denoted
\bea
&&Z = \la 1 \ra = \int d\mu^{}_{V(\alpha M^{\dagger}M)} \nonumber\\
&&\Big{\la} M^{\dagger}_{ab} \dots M^{}_{cd} \Big{\ra} =  \int d\mu^{}_{V(\alpha M^{\dagger}M)} \;   M^{\dagger}_{ab} \dots M^{}_{cd}\; .
\eea
The arguments of a correlation, $M^{\dagger}_{ab} \dots M^{}_{cd}$, are called ``external points''. The partition function and the correlations are evaluated as sums of oriented ribbon graphs with external points build as follows
\begin{itemize}
 \item Vertices (drawn as crossroads of ribbons) are generated by $V(\alpha M^{\dagger}M)$. 
       Each vertex has $2p$ alternating halflines (out of which $p$ are $M^{\dagger}$'s and $p$ 
       are $M$'s) and a weight $N \frac{1}{p} \; g_p \; \alpha^{p}$.
 \item Lines (drawn as ribbons) fall in two categories: internal and external. 
       The internal lines connect a $M^{\dagger}$ 
       halfline and a $M$ halfline. The external lines connect an internal $M^{\dagger}$ 
       (or $M$) halfline with an external $M$ 
       (respectively $M^{\dagger}$) point\footnote{We consider the lines connecting two external 
       points as external.}. A lines has a natural orientation (say from $M^{\dagger}$ to $M$)
       and a weight $\frac{1}{N}$.
 \item Faces are closed circuits formed by the sides of the ribbons. They also fall in two 
       categories, internal and external. The 
       internal faces are closed circuits which {\it do not pass} trough any external point. They 
       have weight $N$. The external faces (containing external points) have weight $1$.
\end{itemize}

\begin{figure}[t]
\begin{centering}
\includegraphics[width=4cm]{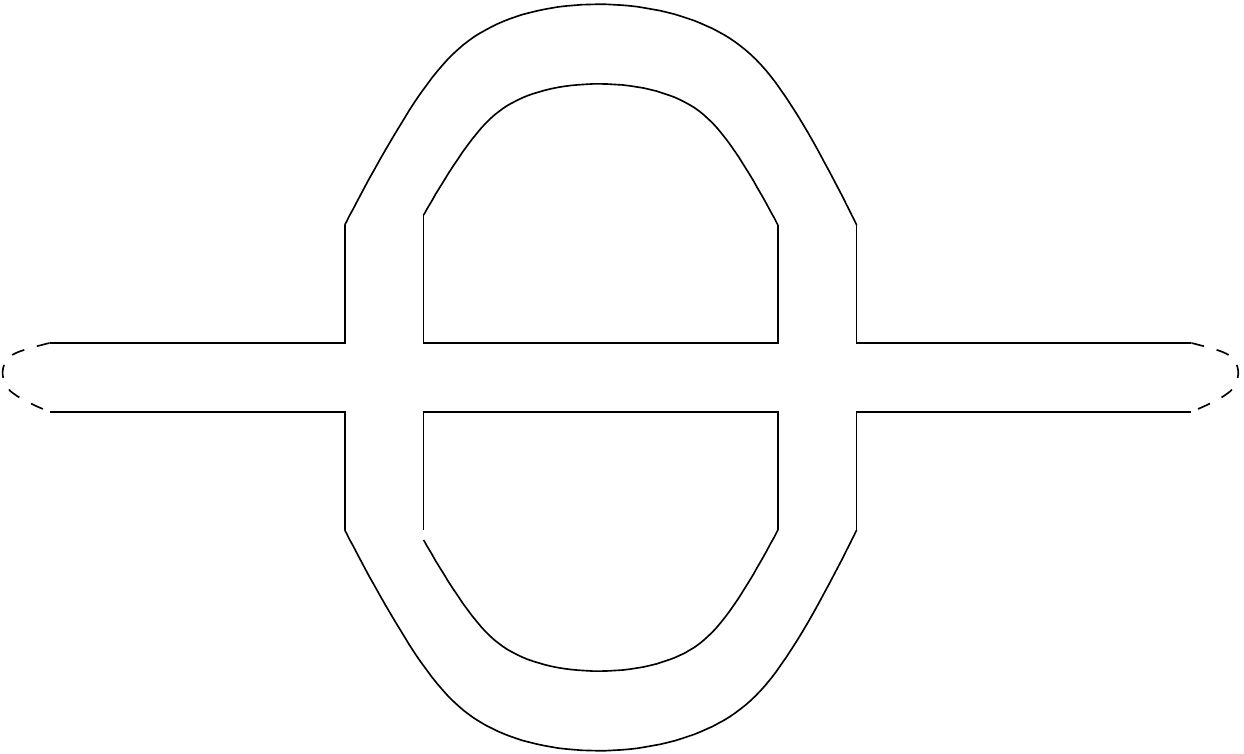}
\caption{A typical ribbon graph.}
\label{fig:sunshine}
\end{centering}
\end{figure}

The identification of the external faces of a graph is slightly non trivial. To count
 the external faces of a graph one ``pinches'' the external points, that is one connects
 the two sides of the ribbon arriving at any external point.

Figure \ref{fig:sunshine} represents a typical ribbon graph. The ``pinching'' of the 
external points is represented by the dotted lines. This graph has two external points, 
two vertices, three internal lines, two external lines two internal faces and 
{\it one} external face.

The partition function $Z=\langle 1 \rangle$ is the sum of graphs with no external points 
(vacuum graphs). The non trivial correlations have the same number of $M$ and $M^{\dagger}$ 
external points. They write 
\bea
\frac{1}{\la 1 \ra} 
\Big{\la} M^{\dagger}_{\mu_1 \nu_1} M^{}_{\nu_1'\mu_1'} \dots M^{\dagger}_{\mu_{N_e}^{} \nu_{N_e}^{} }
 M^{}_{\nu_{N_e}' \mu_{N_e}'}\Big{\ra} =\sum_{\cG} \cA(\cG) \;,
\eea
where $\cG$ are all ribbon graphs with no vacuum connected components (due to the division 
by $\langle 1 \rangle$) and $\cA(\cG)$ is the amplitude of $\cG$. 

To compute $\cA(\cG)$ we denote $V$ the number of vertices (of coordinations 
$2p^{}_1,\dots 2p^{}_V$), $L$ the number of internal lines, $N_e$ the number of 
external points, $F$ the {\it total} number of faces and $B$ the number of external 
faces of $\cG$. We have the two topological relations
\bea
 2\sum_{v}p_v-N_e=2L  \qquad   V-L+F=2-2g \;,
\eea
with $g$ the genus of the graph $\cG$ (thus for the graph of figure \ref{fig:sunshine} $g=0$). Furthermore, for an external face $b$ of $\cG$, we index the $N_b$ external points $M^{\dagger}$ by $i^{(b)}$ and the $N_b$ external points $M$ by $j^{(b)}$. We associate to the face $b$ the ordered set 
\bea
b= \{i^{(b)}_1,j^{(b)}_1,\dots i^{(b)}_{N_b}, j^{(b)}_{N_b} \}  \;,
\eea
and we have
\bea
&& b_1\cap b_2=\emptyset \qquad \forall b_1\neq b_2 \nonumber\\
&&\bigcup_{b} \{i^{(b)}_1\dots i^{(b)}_{N_b}\}= \{1 \dots N_e\} \qquad
\bigcup_{b} \{j^{(b)}_1\dots j^{(b)}_{N_b}\}= \{1 \dots N_e\} \; .
\eea

With these notations the amplitude of $\cG$ writes 
\bea \label{eq:ampli}
 \cA(\cG)&=&K(\cG) \; \alpha^{\sum_{v}p_v} N^{V-(L+N_e)+(F-B)} \prod_{v=1}^{V} g_{p_v}  \nonumber\\
&&\prod_{b} \delta_{ \nu^{}_{i^{(b)}_1} \nu'_{j^{(b)}_1}} \delta_{\mu'_{j^{(b)}_1} \mu^{}_{i^{(b)}_2}} \dots
\delta_{ \nu^{}_{i^{(b)}_{N_b}} \nu'_{j^{(b)}_{N_b}}} \delta_{\mu'_{j^{(b)}_{N_b}} \mu^{}_{i^{(b)}_1}} 
\;,
\eea 
where $K(\cG)$ is some combinatorial coefficient.

Of particular interest in the sequel is the resolvent defined as
\bea
 \omega_N(z) = \frac{1}{N} \frac{\Big{\la} \text{Tr} \Big{(}\frac{1}{z-M^{\dagger}M} \Big{)}\Big{\ra}}{\la 1\ra}=
\sum_{p=0}^{\infty} \frac{1}{z^{p+1}} 
\frac{1}{N} \frac{\Big{\la} \text{Tr} [(M^{\dagger}M)^p] \Big{\ra}}{\langle 1 \rangle} \;.
\eea
We will denote its limit when $N\to\infty$ by $\omega(z)$. The coefficient  
\bea
 d_{p;(N)} =\frac{1}{N} \frac{1}{\langle 1 \rangle} \Big{\la} \text{Tr} [(M^{\dagger}M)^p] \Big{\ra} \;,
\eea
is the sum over all connected vacuum graphs having a special vertex of coordination $2p$ and 
weight $1$ (the trace and matrix product over the external points allows us to reinterpret 
them as the halflines of this special vertex). The global power in 
$N$ of a graph contributing to $d_{p;(N)}$ is, using  eq. (\ref{eq:ampli}),
\bea
\frac{1}{N} N^{(V-1)-L+F} = N^{-2g} \; ,
\eea
and we recognize the classical result that $d_{p}=\lim_{N\to \infty} d_{p;(N)}$ is the sum over connected {\it planar} vacuum graphs with 
a special vertex of weight $1$ and coordination $2p$. 

Before concluding this section we present the relation between the connected planar one vertex functions $d_p$ and the connected amputated planar 
functions with one external face. Our motivation is twofold: first, from a quatum field theory perspective, the latter 
dominate the floating Wilsonian action\footnote{To be distingushed from the effective action which is generated 
by one particle irreducible graphs.}\cite{Samhof} and second, the reasoning we present below is similar to the one we 
will use in section \ref{sec:counting} to derive our fundamental equation.

A planar connected amputated function with one external face and $2t$ external points writes
\bea
N^{2t} \Big{\la} M^{\dagger}_{a^{}_1b^{}_1}M^{}_{b_1'a_2'}\dots M^{}_{b_t'a_t'}\Big{\ra}^{g=0,B=1}
=G_t \; \delta_{b^{}_1b_1'} \delta_{a_2'a^{}_2}\dots \delta_{b^{}_tb_t'} \delta_{a_t'a^{}_1}  \; .
\eea

\begin{figure}[htb]
\begin{centering}
\includegraphics[width=5cm]{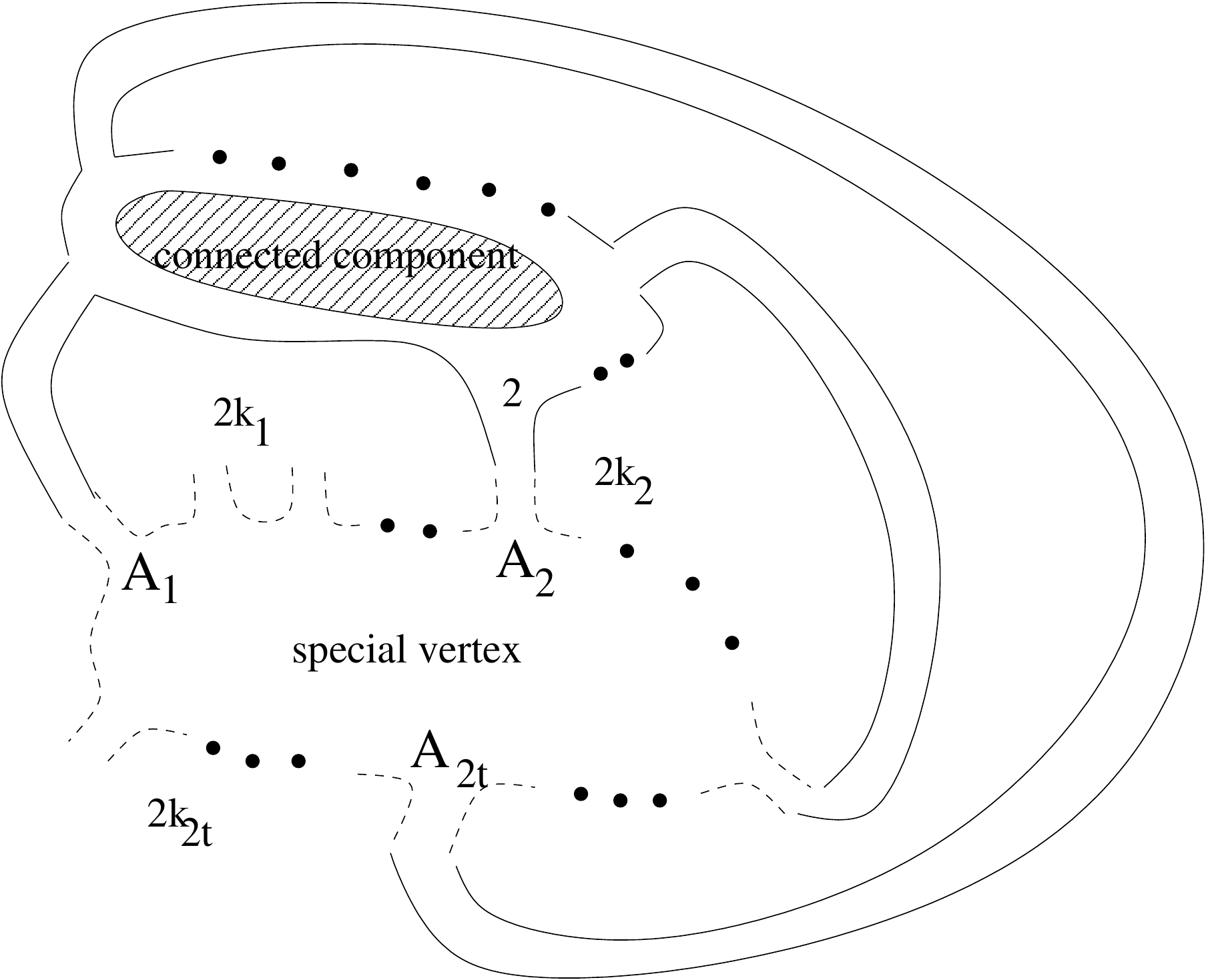}
\caption{A graph $\cG$ contributing to $d_p$.}
\label{fig:omegcon}
\end{centering}
\end{figure}

Consider a graph $\cG$ contributing to $d_p$ (depicted in figure \ref{fig:omegcon}). $\cG$ is a connected planar vacuum graph with a special vertex of 
coordination $2p$ and weight $1$. The special vertex is represented by dashed lines in figure \ref{fig:omegcon}. Consider a halfline of the special vertex, denoted $A_1$ in figure \ref{fig:omegcon}. As $\cG$ is planar, if we erase the special vertex $A_1$ becomes an external point of some planar connected component ${\cal C}$ with one external face, represented as shaded in figure \ref{fig:omegcon}. Say the ${\cal C}$ has $2t$ external points.
All of them originate from halflines on the special vertex, $A_1, A_2,\dots A_{2t}$. We denote $2k_1$ the number of halflines on the special vertex between $A_1$ and $A_2$, $2k_2$ the number of halflines between $A_2$ and $A_3$ etc. up to $2k_{2t}$ the number of halflines between 
$A_{2t}$ and $A_1$. As $\cG$ is planar, the halflines $2k_1$ must connect among themselves and therefore form a graph corresponding 
to $d_{k_1}$. The same is true for $2k_2$, $2k_3$ etc. Therefore $d_p$ writes 
\bea\label{eq:dpg}
 d_{p}=\sum_{t=1}^{p} G_t \sum_{\stackrel{k_1\dots k_{2t}=0}{k_1+\dots k_{2t}=p-t}}^{p-1} d_{k_1} 
\dots d_{k_{2t}} \; .
\eea

Equations (\ref{eq:dpg}) can be solved iteratively for $G_t$. The first equations are
\bea
 d_1&=&G_1 \nonumber\\
 d_2&=&G_2+ 2 G_1  d_1 \nonumber\\
 d_3&=&G_3+ 4 G_2 d_1+ G_1 (d_1^2+2d_2) \; ,
\eea
which are solved by
\bea
 G_1 &=& d_1 \nonumber\\
 G_2 &=& d_2-2G_1d_1=d_2-2d_1^2 \nonumber\\
 G_3 &=& d_3-4 G_2 d_1 - G_1 (d_1^2+2d_2) = d_3 - 6d_1d_2 +7 d_1^3 \; .
\eea

Before concluding this section we introduce a notation. As $V(x)=-\sum_{p}\frac{1}{p}g_p x^p$ is a formal power series, we will denote its derivative (in the sense of formal power series) by 
\bea\label{eq:V'}
V'(x) = -\sum g_p x^{p-1} \; .
\eea 

\section{The Diagrammatic Equation}\label{sec:counting}

In quantum field theory the Polchinski equation \cite{Polchinski:1983gv} is an equation for the floating Wilsonian action $S$ at scale $\Lambda$ and writes 
\bea\label{eq:POL}
 \partial_{\Lambda} S = \frac{1}{2} \int_{x,y} \frac{\delta S}{\delta \phi(x)} 
   \partial_{\Lambda} K(x,y) \frac{\delta S}{\delta \phi(y)} 
   -\frac{1}{2} \int_{x,y} \frac{\delta^2 S}{\delta \phi(x) \delta \phi(y)} 
   \partial_{\Lambda} K(x,y) \; ,
\eea
where $K(x,y)$ is the propagator with ultraviolet cutoff $\Lambda$. The action $S$ is the sum of connected amputated graphs, and the propagator $K$ is associated to lines.  This equation naturally translates in a graph classification: a line in a graph contributing to the action $S$ can either be a ``tree line'' separating two distinct effective vertices (the first term in eq. (\ref{eq:POL})) or it can be a ``loop line'' for some effective vertex (the second term in eq. (\ref{eq:POL})). Adapting this idea to ribbon graphs will lead to our fundamental equation. There are two important aspects one needs to take into account. First, in stead of dealing with the floating action, we will deal with the resolvent of the matrix model. Second, we will look for an equation valid in the $N\to\infty$ limit, thus involving only planar graphs.

\begin{figure}[htb]
\begin{centering}
\includegraphics[width=4cm]{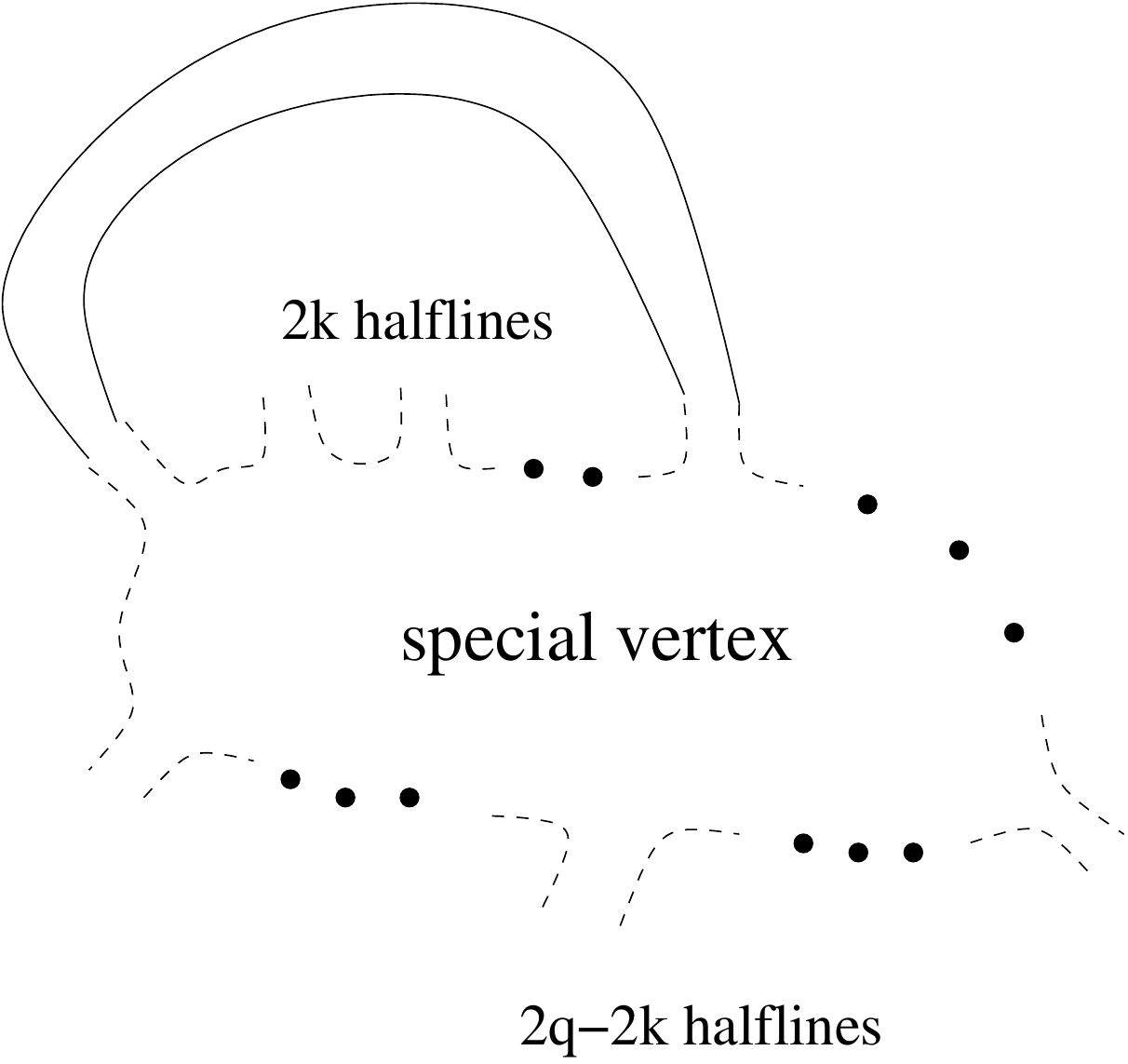}
\caption{A loop line on the special vertex.}
\label{fig:gaus}
\end{centering}
\end{figure}

First consider a free model, obtained by setting $\alpha=0$ in eq. (\ref{eq:definition}). We will denote (the $N\to\infty$ limit of) the resolvent of the free model $\omega^{(0)}(z)$ and its coefficients $d^{(0)}_p$.

With this notations, $d^{(0)}_{q+1}$ is the sum of all planar vacuum graphs with one special vertex of coordination $2q+2$ and weight $1$. Consider a line starting on the special vertex. As the graph contains no other vertices, this line must also end on the special vertex. It is therefore allways a loop line for the special vertex. The line encloses a certain number of halflines, say $2k$, and leaves on the exterior $2q-2k$ halflines as in figure \ref{fig:gaus}. As the initial graph is planar, the interior (exterior) halflines necessarily connect into planar graphs with one special vertex of coordination $2k$ (respectively $2q-2k$).
Therefore
\bea \label{eq:gaus}
 d^{(0)}_{q+1}=\sum_{k=0}^{q} d^{(0)}_k d^{(0)}_{q-k}, \qquad d^{(0)}_0=1\; .
\eea
The attentive reader will note that up to now we did nothing new: eq. (\ref{eq:gaus}) is nothing but equation eq. (\ref{eq:dpg}) supplemented by the condition that the connected amputated functions of a free model are trivial, $G_t=\delta_{t1}$. 

\begin{figure}[htb]
\begin{centering}
\includegraphics[width=4cm]{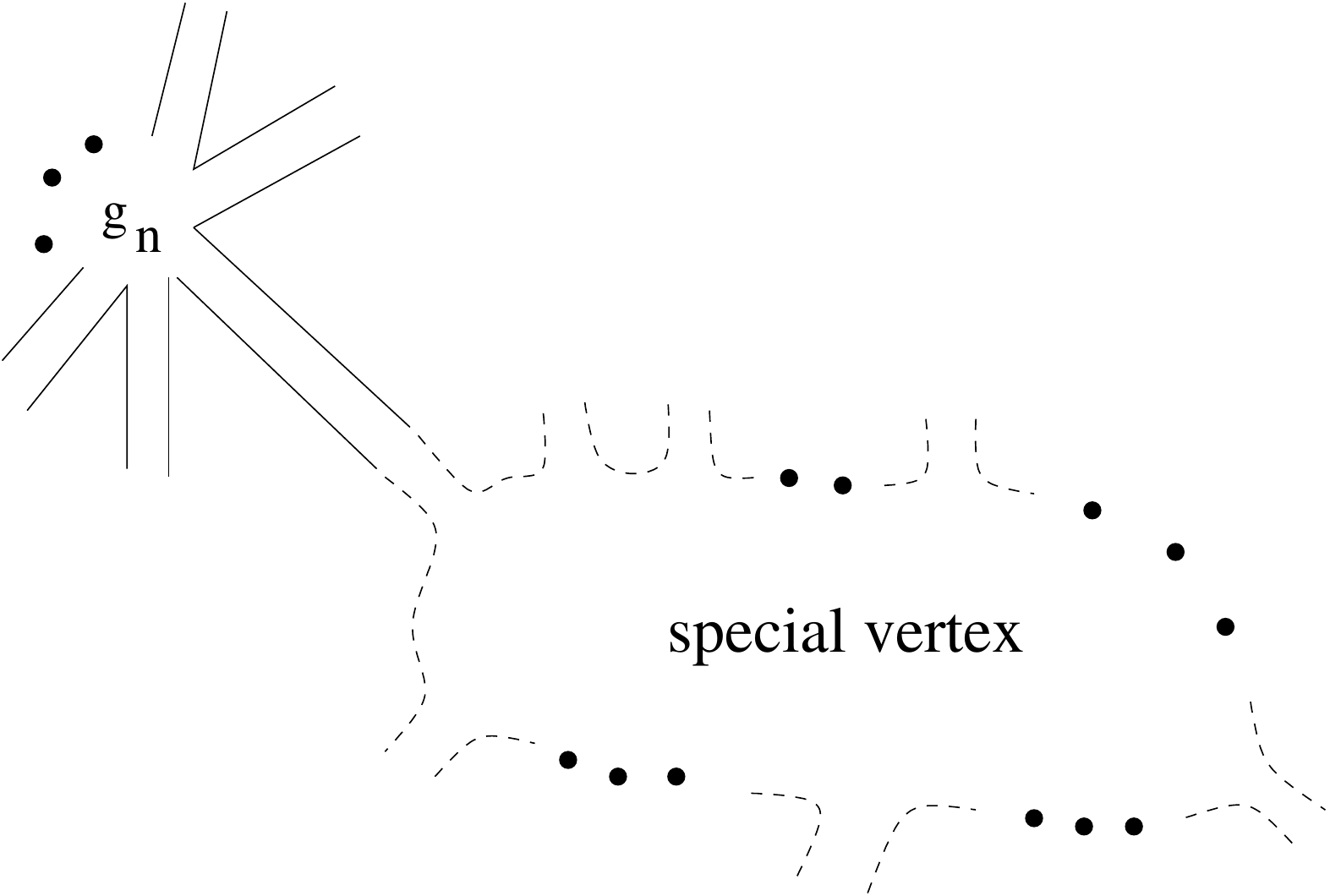}
\caption{A tree line in the interacting model.}
\label{fig:tree}
\end{centering}
\end{figure}

For an interacting model, a line originating on the special vertex of $d_{q+1}$ can either (as for the free model) end on the special vertex, or it can end on a $g_n$ vertex as in figure \ref{fig:tree}. Up to a weight factor, the graphs contributing to the latter are in one to one correspondence with the connected planar vacuum graphs with a special vertex of coordination $2q+2n$. To compute the mismatch in weight factors, recall that the weight of the vertex $g_n$ is $\frac{1}{n}g_n\alpha^n$ and note that the line has a choice of one among the $n$ halflines of appropriate orientation on $g_n$, therefore
\bea\label{eq:fund}
 d_{q+1} = \sum_{k=0}^{q} d_k d_{q-k}+ \sum_{n=1}^{\infty} g_n \alpha^{n}  d_{q+n} \; .
\eea

This are our fundamental diagrammatic equations. The rest of this paper is devoted to their analysis.

\section{The Saddle Point Analysis}\label{sec:saddle}

In this section we derive the equations (\ref{eq:fund}) by an independent method. First, by a carefull saddle point analysis adapted to a square non hermitian random matrix model, we will derive an equation obeyed by the resolvent $\omega(z)$ and then, by developping in powers of $z$, we will rederive the equations (\ref{eq:fund}) for the coefficients $d_p$.

To evaluate an integral with measure (\ref{eq:definition}) by a saddle point method one needs to change variables from $M^*_{ab}, M^{}_{ab}$ to the eigenvalues of $M^{\dagger} M$ \cite{Di Francesco:2002ru}, \cite{Morris:1990cq}. At this stage one has a choice. One can either chose to change variables to $\lambda_i$ the eigenvalues of $M^{\dagger} M$, or to change variables to $y_i$, the eigenvalues of $\sqrt{M^{\dagger} M}$. For example in \cite{Di Francesco:2002ru} the author chooses the former. However, for square matrices, there is a heavy price to be paid in doing this choice.

To understand the problem consider first Euler's gamma function
\bea
 \Gamma(n)=\int_{0}^{\infty} t^{n-1} e^{-t} dt \; .
\eea
It is well known that this integral is governed by a saddle point $t_0=n-1$, and the saddle point evaluation of it proves Stirling's formula. However a polynomial change of variables $u=t^{n}$ yields
\bea
 \Gamma(n)=\frac{1}{n} \int_{0}^{\infty} e^{-u^{\frac{1}{n}}}du \; ,
\eea
and in this new form the integral is {\it not} governed by a saddle point: the Jacobian of the polynomial change of variables absorbed it.

This same phenomenon occurs for square non hermitian matrices. Following 
\cite{Di Francesco:2002ru} one writes (up to some normalization) the partition function
\bea\label{eq:treaba}
 Z= \Big{[}\int_{0}^{\infty} \prod_k d\lambda_k\Big{]} \prod_{i<j}(\lambda_j-\lambda_i)^2 
e^{-N \Big{(} \sum_{i} \lambda_i + \sum_i V(\alpha \lambda_i) \Big{)} } \; .
\eea
For simplicity let us set $\alpha=0$. Raising the Van Der Monde determinant in the exponent yields
\bea
 \sum_{i} \lambda_i-\frac{2}{N}\sum_{j < i} \ln |\lambda_j-\lambda_i| \; .
\eea

The intergal (\ref{eq:treaba}) is not governed by a saddle point: the would be saddle point 
equations 
\bea\label{eq:deriv}
 1 - \frac{2}{N} \sum_{j\neq i} \frac{1}{\lambda_i-\lambda_j}=0, \qquad \forall i
\eea
are incompatible: adding them all up yields $N=0$. Considering, as in \cite{Di Francesco:2002ru}, rectangular matrices solves this problem. In that case the saddle point equations pick up an extra term and are compatible. As previously stated, their results can be applied for square matrices only after taking the large $N$ limit.

In the sequel we will take the second alternative and use the variables $y_i$, eigenvalues of $\sqrt{M^{\dagger}M}$. The Jacobian of this change of variables has been computed in \cite{Morris:1990cq} and the partition function writes (again up to some normalization)
\bea\label{eq:appropriate}
 Z&=&\Big{[}\int_{0}^{\infty}\prod_k  dy_k\Big{]} \prod_{i<j}(y_j^2-y_i^2)^2 (\prod_{i}y_i) 
e^{-N\Big{(}\sum y_i^2+ \sum_i V(\alpha y_i^2)\Big{)} }
\nonumber\\
&=&  \Big{[}\int_{0}^{\infty}\prod_k dy_k \Big{]}   e^{-N F(\{y_i\})}
 \; ,
\eea
where we denoted
\bea
F(\{y_i\})=\sum_i y_i^2+ \sum_i V(\alpha y_i^2) -\frac{1}{N} \sum_i \ln y_i-\frac{2}{N} \sum_{i<j} \ln |y_j^2-y_i^2| \; .
\eea

We have the following lemma

\begin{lemma}\label{lem:omeg}
  $\omega(z)$ obeys the quadratic equation
  \bea\label{eq:finaomega}
\omega(z)^2 - P(z) \omega(z) +Q(z)=0 \; ,
\eea
with
\bea\label{eq:finomega'}
&&P(z)=1+ \alpha V'(\alpha z)  \nonumber\\
&& Q(z)= \frac{1}{z} + \frac{1}{z} \frac{\alpha }{N} 
\sum_i \frac{zV'(\alpha z)-m_i^2 V'(\alpha m_i^2)}{z-m_i^2} \; .
\eea
\end{lemma}
\noindent{\bf Proof:} The proof is a variation on the standard derivation of the quadratic equation obeyed by the resolvent of a hermitian matrix model. 

We start by writing the resolvent as
\bea\label{eq:omeome}
 \omega_N(z)&=&\frac{1}{N}\frac{1}{ \la 1\ra }
\Big{\la} \text{Tr}\Big{(}\frac{1}{z-M^{\dagger}M}\Big{)} \Big{\ra}\nonumber\\
  &=&\frac{1}{N} \frac{  \Big{[}\int_{0}^{\infty} \prod_kdy_k \Big{]} \sum_{i}
   \frac{1}{z-y_i^2}  e^{-NF(y_i)} }{  \Big{[}\int_{0}^{\infty}\prod_k dy_k \Big{]} e^{-NF(y_i)} }\; .
\eea
By slight abuse of notations we denote the leading contribution (in powers of $\frac{1}{N}$) in eq. (\ref{eq:omeome}) also by $\omega_N(z)$.  At large $N$ both the numerator and the denominator above are dominated by the same saddle point $y_i=m_i$, with $m_i$ the solution of the equations 
\bea\label{eq:derivate}
\frac{\partial}{\partial y_i}F=2y_i + \alpha 2y_i V'(\alpha y_i^2)
-\frac{1}{N} \frac{1}{y_i} -\frac{2}{N} \sum_{j\neq i}\frac{ 2y_i }{y_i^2-y_j^2}=0 \; ,
\eea 
where $V'$ is the formal power series in eq. (\ref{eq:V'}). The resolvent (at leading order in $\frac{1}{N}$) is then
\bea
 \omega_N(z)= \frac{1}{N} \sum_{i} \frac{1}{z-m_i^2} \; .
\eea 

Eq. (\ref{eq:appropriate}) ensures that all $m_i$ are strictly positive. In order to find the equation for $\omega_N(z)$ we note that any linear combination of the saddle point equations is zero, hence
\bea\label{eq:saddle}
0&=&\frac{1}{N} \sum_i \frac{1}{z-m_i^2}
\frac{1}{2m_i} \Big{(}\frac{\partial}{\partial y_i} F\Big{)}_{y_i=m_i} \\
&=& \frac{1}{N} \sum_i \frac{1}{z-m_i^2} 
\Big{(}1 +\alpha V'(\alpha m_i^2) - \frac{1}{2N} \frac{1}{m_i^2} 
-\frac{2}{N} \sum_{j\neq i}\frac{ 1 }{m_i^2-m_j^2} \Big{)} \; . \nonumber
\eea
In the case of a hermitian matrix model one employs a similar trick, but without dividing by $2m_i$. Each term in the last line of eq. (\ref{eq:saddle}) receives a different treatment. The first term (also present for hermitian matrices) is identified as 
\bea
 \frac{1}{N} \sum_i \frac{1}{z-m_i^2} = \omega_N(z) \; .
\eea
The second term (again present for hermitian matrices) rewrites by adding and subtracting $\alpha V'(\alpha z)$ as
\bea
 &&\frac{1}{N} \sum_i \frac{1}{z-m_i^2}
 \Big{(}\alpha V'(\alpha m_i^2) -\alpha V'(\alpha z) + \alpha V'(\alpha z) \Big{)} \nonumber\\
 &&=\alpha \; V'(\alpha z) \; \omega_N(z) +\frac{1}{N} \sum_{i} \frac{\alpha V'(\alpha m_i^2) -\alpha 
V'(\alpha z)   }{z-m_i^2} \; .
\eea
The third term receives a similar treatment namely it is rewritten as
\bea\label{eq:new}
-\frac{1}{N^2}\sum_i \frac{1}{z-y_i^2} \Big{(} \frac{1}{y_i^2} - \frac{1}{z} +\frac{1}{z} \Big{)}
= -\frac{1}{2N} \; \frac{1}{z} \omega_N(z) + \frac{1}{2N} \; \frac{1}{z} \omega_N(0) \; .
\eea
This is a new term, not present in the hermitian matrices case.
Finally the last term computes (as for of hermitian matrices) to
\bea
- \frac{2}{N^2} \sum_i \sum_{j\neq i} \frac{1}{m_i^2-m_j^2} \frac{1}{z-m_i^2}  = - \Big{(}\omega_N(z)^2+\frac{1}{N} \frac{d\omega_N(z)}{dz} \Big{)} \; .
\eea
Substituting everything into eq. (\ref{eq:saddle}) yields
\bea\label{eq:aproapeomega}
0&=&\omega_N(z)+ \alpha V'(\alpha z) \; \omega_N(z)+
\frac{\alpha}{N} \sum_i \frac{V'(\alpha m_i^2) -V'(\alpha z) }{z-m_i^2} \nonumber\\
&&-\frac{1}{2N}\frac{1}{z} \omega_N(z) +\frac{1}{2N} \frac{1}{z} \omega_N(0)-
 \Big{(}\omega_N(z)^2+\frac{1}{N} \frac{d\omega_N(z)}{dz} \Big{)} \; .
\eea

The new terms of equation (\ref{eq:new}) prevent us from taking $N\to \infty$ at this stage:  although $\omega_N(0)$ is well defined for any finite $N$, its limit $\omega(0)$ is ill defined hence one can not conclude anything about the limit of $\frac{1}{N}\omega_N(0)$. To get around this problem we use again the saddle point equations (\ref{eq:derivate}) to write
\bea \label{eq:crucial}
0=\frac{1}{N}\sum_i \frac{1}{2m_i} \partial_i F(m_i)= 1+ \frac{\alpha}{N} \sum_i V'(\alpha m_i^2) +\frac{1}{2N} \omega_N(0) \; .
\eea

Substituting eq. (\ref{eq:crucial}) into eq. (\ref{eq:aproapeomega}) yields
\bea
0&=&-\frac{1}{N} \frac{d\omega_N(z)}{dz} - \omega_N(z)^2 + \omega_N(z) -\frac{1}{2N}\frac{1}{z} \omega_N(z)+ \alpha V'(\alpha z) \omega_N(z)\nonumber\\
&& + 
\frac{\alpha}{N} \sum_i \frac{V'(\alpha y_i^2) -V'(\alpha z) }{z-m_i^2}
+\frac{1}{z} \Big{[}- \frac{\alpha}{N} \sum_i V'(\alpha m_i^2) -1 \Big{]} \; ,
\eea
which rewrites after rearranging the terms as
\bea
0&=&- \frac{1}{N} \frac{d\omega_N(z)}{dz} -\frac{1}{2N}\frac{1}{z} \omega_N(z)
- \omega_N(z)^2+ \Big{(}1+ \alpha V'(\alpha z)  \Big{)} \omega_N(z) -\frac{1}{z} \nonumber\\
&& +  \frac{\alpha}{N} \sum_i
\Big{(}  \frac{V'(\alpha m_i^2) -V'(\alpha z) }{z-m_i^2}
-\frac{1}{z} V'(\alpha m_i^2)
 \Big{)} \; .
\eea
In the large $N$ limit the first two terms vanish and a straightforeward computation proves the lemma \ref{lem:omeg}.
\qed 
\\
\bigskip

Recall that $\omega(z)=\sum_{p}\frac{1}{z^{p+1}}d_p$, therefore lemma \ref{lem:omeg} can be translated into an equation for the coefficients $d_p$. In the reminder of this section we will show that this equation is exactly our diagrammatic equation.

The quadratic equation (\ref{eq:finaomega}) for $\omega(z)$ can be written as
\bea\label{eq:prepar}
 \omega(z) - \frac{1}{z} - \omega^2(z) = - \alpha V'(\alpha z)  \omega(z) + 
\frac{1}{z} \frac{\alpha }{N} \sum_i \frac{zV'(\alpha z)-m_i^2 V'(\alpha m_i^2)}{z-m_i^2} \; ,
\eea
and substituting $\omega(z)=\sum_{p=0}z^{-p-1}d_p$, the left hand side becomes
\bea\label{eq:lhs}
&&\sum_{p=0}^{\infty}\frac{1}{z^{p+1}}d_p - \frac{1}{z}  - \sum_{p=0}^{\infty} \sum_{q=0}^{\infty}\frac{1}{z^{p+q+2}} d_p d_q
\nonumber\\
&&= \sum_{q=0}^{\infty} \frac{1}{z^{q+2}}\Big{(}d_{q+1} - \sum_{k=0}^{q} d_kd_{q-k}  \Big{)}
\; .
\eea

The right hands side in eq. (\ref{eq:prepar}) needs a bit more work.
We first express the coefficients $d_p$ in terms of the saddle point solution $m_i$ as
\bea\label{eq:dpmi}
 \omega(z)=\sum_{p=0}^{\infty} \frac{1}{z^{p+1}}d_p = \frac{1}{N} \sum_{i}\frac{1}{z-m_i^2}
\Rightarrow d_p = \frac{1}{N} \sum_i (m_i^2)^p\; ,
\eea
and substituting $V'$ from eq. (\ref{eq:V'}), the second term on the right hand side of (\ref{eq:prepar}) becomes
\bea
&&- \frac{1}{z} \frac{\alpha }{N} \sum_i \sum_{n=1}^{\infty} g_n \alpha^{n-1}
\frac{z^{n}-(m_i^{2})^n}{z-m_i^2} = -\frac{1}{z} \sum_{n=1}^{\infty} g_n \alpha^{n}
\sum_{t=0}^{n-1} z^{n-1-t} \frac{1}{N} \sum_i (m_i^{2})^t \nonumber\\
&&=- \frac{1}{z} \sum_{n=1}^{\infty} g_n \alpha^{n}
\sum_{t=0}^{n-1} z^{n-1-t} d_t\;.
\eea
Using again the definition of $V'$, the right hand side of (\ref{eq:prepar}) becomes
\bea\label{eq:rhs}
&&\alpha \sum_{n=1}^{\infty} g_n \alpha^{n-1} z^{n-1} \sum_{t=0}^{\infty} \frac{d_t}{z^{t+1}}
-  \sum_{n=1}^{\infty} g_n z^{n-1} \alpha^{n}
\sum_{t=0}^{n-1} \frac{d_t}{z^{t+1}} \nonumber\\
&&= \sum_{n=1}^{\infty} g_n z^{n-1} \alpha^{n}
\sum_{t=n}^{\infty} \frac{d_t}{z^{t+1}}
= \sum_{n=1}^{\infty} g_n z^{n-1} \alpha^{n}
\sum_{q=0}^{\infty} \frac{d_{q+n}}{z^{q+n+1}} \nonumber\\
&& = \sum_{q=0}^{\infty} \frac{1}{z^{q+2}} 
\sum_{n=1}^{\infty} g_n  d_{q+n} \alpha^{n} \; .
\eea
Equating the coefficients of $z^{-q-2}$ in eq. (\ref{eq:lhs}) and (\ref{eq:rhs}) yields
\bea
 d_{q+1} = \sum_{k=0}^{q} d_k d_{q-k}+ \sum_{n=1}^{\infty} g_n \alpha^{n}  d_{q+n} \; ,
\eea
which is exactly our diagrammatic equations (\ref{eq:fund}).

A selfconsistency check is provided by the the free model ($\alpha=0$). In this case eq. (\ref{eq:finaomega}) writes
\bea
\omega(z)^2 - \omega(z) +\frac{1}{z}=0 \Rightarrow
\omega(z)=\frac{1}{2}\Big{(}1-\sqrt{1-\frac{4}{z}} \Big{)} \; ,
\eea
as $\omega(z)\sim_{z\to \infty} z^{-1}$. The eigenvalues distribution of the free model reads 
\bea
&&\tilde\rho(\lambda^2)=\frac{1}{2\pi \imath} [\omega(\lambda^2+\imath 0)-\omega(\lambda^2-\imath 0)]= 
\frac{1}{N} \sum\delta(\lambda^2-m_i^2)\nonumber\\
&&= \tilde \rho(\lambda^2)=\frac{1}{2\pi} \frac{1}{\lambda} \sqrt{4-\lambda^2} \; ,
\eea
and we recover Wigner's semicircle law 
\bea
\rho(\lambda) d\lambda = \tilde\rho(\lambda^2) d(\lambda^2) = \frac{1}{\pi}  \sqrt{4-\lambda^2} \;d\lambda \; .
\eea 

\section{The Perturbative Solution}\label{sec:sol}

In this section we will solve our diagrammatic equation in perturbations. We start by the free model, and subsequently work our way to the solution of a generic model.

\begin{lemma}\label{lem:gaus}
 The solution $d_p^{(0)}$ of the free model (equations (\ref{eq:gaus})) is
 \bea
  d^{(0)}_k=\frac{1}{k+1} C^{2k}_k, \quad \forall k \; .
 \eea
\end{lemma}

{\bf Proof:} A direct computation shows that $d^{(0)}_0=1$. We proceed by induction. Suppose that
\bea
d^{(0)}_k=\frac{1}{k+1} C^{2k}_k, \quad \forall k\le q \; .
\eea
Eq. (\ref{eq:gaus}) then reads
\bea\label{eq:supermeserie}
 d^{(0)}_{q+1}&=&\sum_{k=0}^q \frac{1}{k+1}\frac{1}{q-k+1} C^{2k}_k C^{2q-2k}_{q-k}\nonumber\\
&=&\frac{1}{q+2} \sum_{k=0}^q \Big{(} \frac{1}{k+1}+\frac{1}{q-k+1} \Big{)} C^{2k}_k C^{2q-2k}_{q-k} 
\nonumber\\
&=&\frac{1}{q+2} \sum_{k=0}^q \frac{2}{k+1} C^{2k}_k C^{2q-2k}_{q-k} \;,
\eea
where in the second line a change of the dummy variable identifies the two terms. Separating the term $q=k$ in the sum gives
\bea
d^{(0)}_{q+1}&=&\frac{1}{q+2} \Big{(} \frac{2}{q+1}C^{2q}_q+ \sum_{k=0}^{q-1} \frac{2}{k+1} C^{2k}_k C^{2q-2k}_{q-k}  \Big{)} \nonumber
\\
&=&\frac{1}{q+2} \Big{(} \frac{2}{q+1}C^{2q}_q+ \sum_{k=0}^{q-1} \frac{2}{k+1} C^{2k}_k 
 \Big{(} 4-\frac{2}{q-k} \Big{)} C^{2q-2-2k}_{q-1-k}  \Big{)} \; ,
\eea 
which further computes to
\bea
 d^{(0)}_{q+1}=\frac{1}{q+2} &\Big{(}& \frac{2}{q+1}C^{2q}_q+ 4 \sum_{k=0}^{q-1} 
\frac{2}{k+1} C^{2k}_k C^{2q-2-2k}_{q-1-k}  \nonumber\\
  &-& 4 \sum_{k=0}^{q-1} \frac{1}{k+1} C^{2k}_k \frac{1}{(q-1)-k+1} C^{2q-2-2k}_{q-1-k}  \Big{)} 
\; .
\nonumber
\eea
Taking into account the first and third lines in eq. (\ref{eq:supermeserie}) allows us to write
\bea
 d^{(0)}_{q+1} &=&\frac{1}{q+2}\Big{(} \frac{2}{q+1}C^{2q}_q+ 4(q+1)d^{(0)}_q
-4 d^{(0)}_{q}\Big{)} \nonumber\\
&=& \frac{1}{q+2}\Big{(} \frac{2}{q+1}C^{2q}_q+ 4\frac{q}{q+1}C^{2q}_q\Big{)} 
=\frac{1}{q+2}C^{2q+2}_{q+1}
\; .
\eea 
\qed
\\
\bigskip

We now turn our attention to the interacting model. We will solve the equation (\ref{eq:fund}) in perturbations around the $\alpha=0$ solution we just derived. We substitute
\bea
d_{q}=\sum_i d_q^{(i)} \alpha^i , \qquad d_q^{(0)} = \frac{1}{q+1}C^{2q}_q, \qquad  d_0^{(i)}=0, i\ge1
\; ,
\eea
in eq. (\ref{eq:fund}) to get
\bea \label{eq:subst}
 \sum_{s=0}^{\infty} d_{q+1}^{(s)} \alpha^s = 
\sum_{k=0}^{q} \sum_{s=0}^{\infty}   \Big{(} \sum_{i=0}^s d_{k}^{(i)} d_{q-k}^{(s-i)} \Big{)}\alpha^s
+\sum_{s=1}^{\infty} \sum_{n=1}^{s} g_n d_{q+n}^{(s-n)} \alpha^s \; .
\eea

Equating the powers of $s$ on the left hand side and right hand side yields  
\bea\label{eq:perturb}
d_{q+1}^{(s)} = \sum_{k=0}^{q} \sum_{i=0}^s d_{k}^{(i)} d_{q-k}^{(s-i)} +
\sum_{n=1}^{s} g_n d_{q+n}^{(s-n)}  \qquad \forall s \ge 1 \; .
\eea

This equation mixes all orders of perturbation from $(0)$ to $(s)$. Solving it comes to write the terms at order $(s)$ as functions of terms at lower orders. To this end we define
\bea\label{eq:Tdef}
&& T^{(1)}_{q+1}=g_1 d^{(0)}_{q+1}\nonumber\\
&& T^{(s)}_{q+1}=\sum_{n=1}^{s} g_n d_{q+n}^{(s-n)} + \sum_{k=0}^{q} \sum_{i=1}^{s-1} d_{k}^{(i)} d_{q-k}^{(s-i)}  ,\; s\ge 2 \; ,
\eea
such that at the order $(s)$, $T^{(s)}$ depends only on lower orders $d^{(0)}$ to $d^{(s-1)}$. 

\begin{lemma}\label{lem:gen}
 The solution of (\ref{eq:perturb}) at the order of perturbations $(s)$ is
\bea
 d_{q+1}^{(s)} =\sum_{p=0}^{q}C^{2q-2p}_{q-p} T^{(s)}_{p+1} \; .
\eea
\end{lemma}
\noindent{\bf Proof:} The proof of this lemma is somewhat technical. We start by inserting the definition of $T^{(s)}$ in eq. (\ref{eq:perturb}) to get (for $s\ge 1$) 
\bea\label{eq:ddet}
 d_{q+1}^{(s)} = T^{(s)}_{q+1} + 2\sum_{k=0}^{q} d^{(0)}_{q-k} d^{(s)}_{k} \; .
\eea

We will solve this equation as a recurion over $q$. However note that the right hand side involves all $d^{(s)}_k$ with $k<q+1$. We will first derive $d^{(s)}_{q+1}$ as a function of only $d^{(s)}_q$ and $T^{(s)}$. Then we will solve this recursion and find $d^{(s)}_{q+1}$ as a function of only $T^{(s)}$. 

{\bf Step 1:} Recalling that $d^{(s)}_{0}=0$ for $s\ge 1$, we have
\bea
 d_{q+1}^{(s)} = T^{(s)}_{q+1} + 2\sum_{k=1}^{q} d^{(0)}_{q-k} d^{(s)}_{k}
=T^{(s)}_{q+1} + 2\sum_{p=0}^{q-1} d^{(0)}_{q-1-p} d^{(s)}_{p+1} \; .
\eea
Reinserting eq. (\ref{eq:ddet}) yields
\bea
d_{q+1}^{(s)} &=& T^{(s)}_{q+1} + 2\sum_{p=0}^{q-1} d^{(0)}_{q-1-p} 
\Big{(} T^{(s)}_{p+1} + 2\sum_{t=0}^{p} d^{(0)}_{p-t} d^{(s)}_{t} \Big{)}
\nonumber\\
 &=& T^{(s)}_{q+1} + 2\sum_{p=0}^{q-1} d^{(0)}_{q-1-p}  T^{(s)}_{p+1} 
+4 \sum_{p=0}^{q-1} \sum_{t=0}^{p} d^{(0)}_{q-1-p} d^{(0)}_{p-t} d^{(s)}_{t} 
\nonumber\\
 &=& T^{(s)}_{q+1} + 2\sum_{k=1}^{q} d^{(0)}_{q-k}  T^{(s)}_{k} 
+4 \sum_{t=0}^{q-1} \sum_{p=t}^{q-1} d^{(0)}_{q-1-p} d^{(0)}_{p-t} d^{(s)}_{t} 
\nonumber\\
&=& T^{(s)}_{q+1} + 2\sum_{k=1}^{q} d^{(0)}_{q-k}  T^{(s)}_{k} 
+4 \sum_{t=0}^{q-1} \sum_{u=0}^{q-t-1} d^{(0)}_{q-1-t-u} d^{(0)}_{u} d^{(s)}_{t} \; .
\eea
The sum over $u$ reproduces the right hand side of eq. (\ref{eq:gaus}), thus
\bea
d_{q+1}^{(s)}&=&T^{(s)}_{q+1} + 2\sum_{k=1}^{q} d^{(0)}_{q-k}  T^{(s)}_{k} 
+4 \sum_{t=0}^{q-1} d^{(0)}_{q-t} d^{(s)}_{t} \; .
\eea

Using $d^{(0)}_0=1$ and again eq. (\ref{eq:ddet}), we have
\bea
d_{q+1}^{(s)}&=&T^{(s)}_{q+1} + 2\sum_{k=1}^{q} d^{(0)}_{q-k}  T^{(s)}_{k} 
+4 \Big{(} \sum_{t=0}^{q} d^{(0)}_{q-t} d^{(s)}_{t} - d^{(s)}_{q}\Big{)}
\nonumber\\
&=&T^{(s)}_{q+1} + 2\sum_{k=1}^{q} d^{(0)}_{q-k}  T^{(s)}_{k} 
+4 \Big{(}\frac{d^{(s)}_{q+1}-T^{(s)}_{q+1}}{2} - d^{(s)}_{q}\Big{)} \; ,
\eea
and regrouping $d^{(s)}_{q+1}$ on the left hand side yields
\bea\label{eq:dqdq}
d_{q+1}^{(s)}= 4 d^{(s)}_{q}- 2 \sum_{k=1}^{q} d^{(0)}_{q-k}  T^{(s)}_{k} +
T^{(s)}_{q+1} \; .
\eea
Our first objective is now achieved, as $d^{(s)}_{q+1}$ is expressed solely in terms of $d^{(s)}_q$ and $T^{(s)}$.

{\bf Step 2:} Reinserting eq. (\ref{eq:dqdq}) into itself $p$ times we find 
\bea
 d_{q+1}^{(s)} &=& 4^{p+1} d^{(s)}_{q-p}-\sum_{k=1}^{q-p}\Big{(} 
\sum_{s=0}^{p}2\cdot 4^s d^{(0)}_{q-k-s} \Big{)} T^{(s)}_{k}\nonumber\\
&+& \sum_{k=q-p+1}^{q}\Big{(}4^{q+1-k}- \sum_{s=0}^{q-k}2\cdot 4^{s} d^{(0)}_{q-k-s}  \Big{)} T^{(s)}_{k} +  T^{(s)}_{q+1} \; .
\eea
Using $d^{(s)}_1=T^{(s)}_1$, setting $p=q-1$ and grouping the first two terms yields
\bea\label{eq:mizerie}
d_{q+1}^{(s)} = \sum_{k=1}^{q}\Big{(}4^{q+1-k}- \sum_{s=0}^{q-k}2\cdot 4^{s} d^{(0)}_{q-k-s}  \Big{)} T^{(s)}_{k} +  T^{(s)}_{q+1} \; .
\eea

By now we have expressed $d^{(s)}$ only in terms of $T^{(s)}$. To simplify this expression, we use the result of appendix \ref{app:1} 
\bea
S_{q}=\sum_{k=0}^{q} \frac{1}{4^{k+1}} d^{(0)}_k = \frac{1}{2}
 \Big{(}1- \frac{1}{4^{q+1}} C^{2q+2}_{q+1} \Big{)} \; ,
\eea
and write 
\bea
 d_{q+1}^{(s)} 
&=& \sum_{k=1}^{q}\Big{(}4^{q+1-k}- 2\cdot 4^{q-k+1} S_{q-k}\Big{)} 
T^{(s)}_{k} +  T^{(s)}_{q+1} \\
&=& \sum_{k=1}^{q}\Big{[}4^{q+1-k}- 2\cdot 4^{q-k+1} 
\frac{1}{2} \Big{(}1-C^{2q-2k+2}_{q-k+1}\frac{1}{4^{q-k+1}} \Big{)}
\Big{]} T^{(s)}_{k} +  T^{(s)}_{q+1}\; , \nonumber
\eea
and finally 
\bea
 d_{q+1}^{(s)} = \sum_{k=1}^{q}C^{2q-2k+2}_{q-k+1} T^{(s)}_{k} +  T^{(s)}_{q+1}
=\sum_{p=0}^{q}C^{2q-2p}_{q-p} T^{(s)}_{p+1} \; .
\eea
\qed
\\
\bigskip

For practical computations it is usefull to recast the perturbative solution of lemma \ref{lem:gen} in terms of generating functions.

\subsection{Generating Functions}

Let us define the generating functions at order $(s)$ 
\bea
\omega^{(s)}(z) = \sum_{q=0}^{\infty} \frac{1}{z^{q+1}} d^{(s)}_q\; , \qquad
T^{(s)}(z)= \sum_{q=0}^{\infty} \frac{1}{z^{q+1}} T^{(s)}_q \;.
\eea
where we set $T^{(s)}_0=0$ for all $s\ge 1$. The solution of the non interacting model of 
lemma \ref{lem:gaus} gives us the generating function at order zero 
\bea\label{eq:omzer}
 \omega^{(0)}(z) = \sum_{q=0} \frac{1}{z^{q+1}} \frac{1}{q+1}C^{2q}_q= 
\frac{1}{2}\Big{(}1- \sqrt{1-\frac{4}{z}}\Big{)}\; .
\eea

For all $s\ge 1$ $d^{(s)}_0=0$, thus lemma \ref{lem:gen} translates for the generating functions at order of perturbations (s) as
\bea\label{eq:os}
 \omega^{(s)}(z)&=&\sum_{q=0}^{\infty} \frac{1}{z^{q+1}}d^{(s)}_{q} = 
\sum_{q=0}^{\infty} \frac{1}{z^{q+2}}d^{(s)}_{q+1}
=\sum_{q=0}^{\infty}  \frac{1}{z^{q+2}} \sum_{p=0}^{q}C^{2q-2p}_{q-p} T^{(s)}_{p+1} \nonumber\\
&=& \sum_{p=0}^{\infty} \frac{1}{z^{p+2}}T^{(s)}_{p+1}  \sum_{q=p}^{\infty} C^{2q-2p}_{q-p} 
\frac{1}{z^{q-p}}= \frac{1}{\sqrt{1-\frac{4}{z}}} T^{(s)}(z) \; .
\eea
The last ingredient one needs is to translate the definition of $T^{(s)}_{q+1}$  (eq. (\ref{eq:Tdef})) for generating functions. We first treat the case $s=1$ 
\bea\label{eq:T1}
 T^{(1)}(z) &=& \sum_{q=0}^{\infty} \frac{1}{z^{q+1}} T^{(1)}_q = \sum_{q=0}^{\infty}
\frac{1}{z^{q+2}} T^{(1)}_{q+1} \nonumber\\
 &=& g_1 \sum_{q=0}^{\infty} \frac{1}{z^{q+2}} d^{(0)}_{q+1} = 
g_1 \Big{(}\omega^0(z)-\frac{1}{z} \Big{)} \;.
\eea
Then, for all $s\ge 2$ we have
\bea\label{eq:Ts}
T^{(s)}(z) &=& \sum_{q=0} \frac{1}{z^{q+2}} T^{(s)}_{q+1}=\sum_{q=0}^{\infty} \frac{1}{z^{q+2}}
\Big{(}\sum_{n=1}^{s} g_n d_{q+n}^{(s-n)} + \sum_{k=0}^{q} \sum_{i=1}^{s-1} d_{k}^{(i)} d_{q-k}^{(s-i)} \Big{)} \nonumber\\
&=& \sum_{n=1}^{s} z^{n-1} g_n \sum_{q=0}^{\infty} d^{(s-n)}_{q+n} \frac{1}{z^{q+n+1}} +
\sum_{i=1}^{s-1} \sum_{q=0}^{\infty} \frac{1}{z^{q+2}} \sum_{k=0}^{q} d_{k}^{(i)} d_{q-k}^{(s-i)} \\
&=& \sum_{n=1}^{s} z^{n-1} g_n \Big{(} \omega^{(s-n)}(z) - 
\sum_{q=0}^{n-1} \frac{1}{z^{q+1}} d^{(s-n)}_{q} \Big{)}
+\sum_{i=1}^{s-1} \omega^{(i)}(z) \omega^{(s-i)}(z) \nonumber \; .
\eea

The equations (\ref{eq:omzer}), (\ref{eq:os}), (\ref{eq:T1}) and (\ref{eq:Ts}) allow one to compute the generating functions $\omega^{(s)}$ order by order in $(s)$. In appendix \ref{app:2} we use them to compute $\omega^{(1)}$, $\omega^{(2)}$ and $\omega^{(3)}$.

\section{Dualities}\label{sec:dualities}

In this last section we will exhibit two duality properties of the perturbative solution. In fact $d_p$ is a power series not only in $\alpha$ but also in $g_k$, that is it writes 
\bea
 d_p&=& \sum_{n_1=0}^{\infty} \sum_{n_2=0}^{\infty} \dots \sum_{n_k=0}^{\infty} \dots \alpha^{(n_1+2n_2+\dots+k n_k+\dots)}
g_1^{n_1} g_2^{n_2}\dots g_{k}^{n_k} \dots \nonumber\\
&&d_{p;g_1^{n_1}g_2^{n_2}\dots g_{k}^{n_k}\dots}^{(n_1+2n_2+\dots+k n_k+\dots)} \; ,
\eea
and the coefficient $d_{p;g_1^{n_1}\dots g_{k}^{n_k}\dots}^{(n_1+\dots+k n_k+\dots)}$ is the number of unlabeled, planar, orientable graphs having a special vertex of coordination $2p$ and $n_1$ vertices $g_1$, $n_2$ vertices $g_2$ and so on and so fort.  

It is clear intuitively that these coefficients are not independent. First, the number of unlabeled graphs with a choice of vertices
is a combinatorial quantity and does not depend on which of the vertices is the  ``special vertex''.
Second, the vertices of weight $g_1$ have coordination two thus they can be viewed as decorations
on the lines of a simpler graph. We present below the precise duality properties of the coefficients $d_{p;g_1^{n_1}\dots g_{k}^{n_k}\dots}^{(n_1+\dots+k n_k+\dots)}$ mirroring these two observations. To this end we start from the definition of $d_p$
\bea
 d_{p} =\frac{1}{N} \frac{1}{\langle 1 \rangle} \Big{\la} \text{Tr} [(M^{\dagger}M)^p] \Big{\ra} 
\; ,
\eea
and denoting $\la \dots \ra_0$ the gaussian ($\alpha=0$) correlation, $d_p$ writes
\bea
d_p=\frac{1}{N} \frac{1}{\langle 1 \rangle} \Big{\la} \text{Tr} [(M^{\dagger}M)^p] 
e^{N\sum_{q=1}^{\infty}\frac{g_q\alpha^q}{q}\text{Tr}[(M^{\dagger}M)^q]}\Big{\ra}_0^{g=0} \; ,
\eea
where the superscript indicates that only planar graphs contribute to the Gaussian average. In the sequel, we only count the planar contributions in all Gaussian averages, but, to simplify notations, we drop the superscript. A perturbative development in all the coupling constants $g_k$ yields
\bea
d_p&=&\frac{1}{N} \frac{1}{\langle 1 \rangle} \sum_{n_1,\dots n_k, \dots=0}^{\infty} N^{n_1+\dots n_k+\dots} \alpha^{n_1+\dots+kn_k+\dots} \nonumber\\
&&\frac{1}{n_1!}\Big{(}\frac{g_1}{1} \Big{)}^{n_1} \dots \frac{1}{n_k!}\Big{(}\frac{g_k}{k} \Big{)}^{n_k} \dots \nonumber\\
&&\Big{\la} \text{Tr} [(M^{\dagger}M)^p] \text{Tr}[(M^{\dagger}M)]^{n_1}\dots \text{Tr}[(M^{\dagger}M)^k]^{n_k}\dots \Big{\ra}_0 \; ,
\eea
which allows us to identify 
\bea\label{eq:dddd}
 &&d^{(n_1+2n_2+\dots k n_k+\dots )}_{p; g_{1}^{n_1}g_{2}^{n_2}\dots g_{k}^{n_k}\dots} = \frac{1}{N} 
\frac{1}{\langle 1 \rangle} 
\frac{N^{n_1+\dots n_k+\dots}}{n_1!\dots n_k!\dots} \; \; \frac{1}{1^{n_1}2^{n_2}\dots k^{n_k} \dots } \nonumber\\
&&\times \Big{\la} \text{Tr} [(M^{\dagger}M)^p] \text{Tr}[(M^{\dagger}M)]^{n_1}\dots \text{Tr}[(M^{\dagger}M)^k]^{n_k}\dots \Big{\ra}_0 \; .
\eea
In the gaussian average above the special vertex does not play any distinguished role,
\bea
&&\Big{\la} \text{Tr} [(M^{\dagger}M)^p] \text{Tr}[(M^{\dagger}M)]^{n_1}\dots \text{Tr}[(M^{\dagger}M)^k]^{n_k}\dots \Big{\ra}_0
\nonumber\\
&&=\Big{\la}  \text{Tr}[(M^{\dagger}M)]^{n_1}\dots \text{Tr} [(M^{\dagger}M)^p]^{n_p+1}\dots \text{Tr}[(M^{\dagger}M)^k]^{n_k}\dots \Big{\ra}_0 \; .
\eea
The only thing which distinguishes between the special vertex and any other vertex is the combinatorial weight. Balancing this combinatorial proves the first duality property
\bea\label{eq:dual1}
\frac{1}{p} \frac{1}{n_p+1} d^{(n_1+\dots p n_p+\dots q (n_q+1)+\dots )}_{p; g_{1}^{n_1}\dots g_{p}^{n_p}\dots g_{q}^{n_q+1}\dots} 
=\frac{1}{q} \frac{1}{n_q+1} d^{(n_1+\dots p (n_p+1)+\dots q n_q +\dots)}_{q; g_{1}^{n_1}\dots g_{p}^{n_p+1}\dots g_{q}^{n_q}\dots} \; ,
\eea
for all $q$ with $n_q\ge 0$. 

To prove the second duality we introduce the modified Gaussian measure
\bea
\int \Big{[} \prod_{ab} dM_{ab} \; dM^*_{ab} \Big{]} \; e^{-xN\Big{(} \text{Tr}(M^{\dagger}M) \Big{)}}
\eea
and we denote the correlations of this measure $\la \dots \ra_{0;x}$. With the help of the modified measure we can write
\bea\label{eq:last}
d^{(n_1+1+\dots k n_k+\dots )}_{p; g_{1}^{n_1+1}\dots g_{k}^{n_k}\dots} = \frac{1}{N} \frac{1}{\langle 1 \rangle} 
\frac{N^{n_1+1+\dots n_k+\dots}}{(n_1+1)!\dots n_k!\dots}  \; \; \frac{1}{1^{n_1+1}2^{n_2}\dots k^{n_k} \dots} \nonumber\\
\Big{\la} \text{Tr} [(M^{\dagger}M)^p] \text{Tr}[(M^{\dagger}M)]^{n_1+1}\dots \text{Tr}[(M^{\dagger}M)^k]^{n_k}\dots \Big{\ra}_0
\nonumber\\
=\frac{1}{N} \frac{1}{\langle 1 \rangle} 
\frac{N^{n_1+1+\dots n_k+\dots}}{(n_1+1)!\dots n_k!\dots}  \;\;
\frac{1}{1^{n_1+1}2^{n_2}\dots k^{n_k} \dots}
\nonumber\\
\Big{(}-\frac{1}{N}\frac{d}{dx}
\Big{\la} \text{Tr} [(M^{\dagger}M)^p] \text{Tr}[(M^{\dagger}M)]^{n_1}\dots \text{Tr}[(M^{\dagger}M)^k]^{n_k}\dots \Big{\ra}_{0,x} \Big{)}\Big{\vert}_{x=1} \; ,
\eea
as the derivative acting on the gaussian measure produces an extra $\text{Tr}(M^{\dagger}M)$ insertion. But the correlations $\la \dots \ra_{0;x}$ of the modified gaussian measure are sums over the same graphs as the usual gaussian correlations $\la \dots \ra_{0}$, with the only difference that the lines of the graphs have a weight $\frac{1}{x N}$ in stead of $\frac{1}{N}$. All graphs contributing to 
the averages in (\ref{eq:last}) have the same number of lines, namely $p+\sum_{j=1}^{\infty} jn_j$.  Hence 
\bea
&&d^{(n_1+1+\dots k n_k+\dots )}_{p; g_{1}^{n_1+1}\dots g_{k}^{n_k}\dots} =
\frac{1}{N} \frac{1}{\langle 1 \rangle} 
\frac{N^{n_1+1+\dots n_k+\dots}}{(n_1+1)!\dots n_k!\dots}  \; \; \frac{1}{1^{n_1+1}2^{n_2}\dots k^{n_k} \dots} \nonumber\\
&&\qquad\Big{(}-\frac{1}{N}\frac{d}{dx}\Big{)}\Big{\vert}_{x=1}(x)^{-(p+\sum_{j=1}^{\infty} jn_j)} \nonumber\\
&&\qquad\Big{\la} \text{Tr} [(M^{\dagger}M)^p] \text{Tr}[(M^{\dagger}M)]^{n_1}\dots \text{Tr}[(M^{\dagger}M)^k]^{n_k}\dots \Big{\ra}_{0} \; .
\eea
Tracking again the weight factors one has
\bea\label{eq:dual2}
 d^{(n_1+1+2n_2+\dots +k n_k+\dots)}_{p; g_{1}^{n_1+1}\dots g_{k}^{n_k}\dots} = 
\frac{\sum_{j=1}^{\infty} j n_j+p}{n_1+1}  \;
d^{(n_1+2n_2+\dots k n_k+\dots)}_{p;g_1^{n_1} g_{2}^{n_2}\dots g_{k}^{n_k}\dots} \; .
\eea

These dualities are very usefull in concrete computations. Using the results of appendix \ref{app:2}, $d_p$ writes up to third order in $\alpha$ as
\bea
 d_q&=& \frac{1}{q+1}C^{2q}_q +\alpha g_1 \frac{q}{q+1}C^{2q}_q + 
\alpha^{2} g_1^2\frac{q}{2} C^{2q}_{q} +\alpha^2 g_2 \frac{3q}{q+2} C^{2q}_{q} \nonumber\\
&+& \alpha^{3} g_1^3 \frac{q(q+2)}{6} C^{2q}_q+\alpha^3 g_1 g_2 \; 3q C^{2q}_q +
\alpha^3 g_3 \frac{10q}{q+3}C^{2q}_q \; .
\eea
The first duality (\ref{eq:dual1}) implies for instance
\bea
 \frac{1}{2} d^{(3)}_{2,g_3}= \frac{1}{3}d^{(2)}_{3,g_2}\quad 
  d_{1,g_3}^{(3)}=\frac{1}{3}d_{3,g_1}^{(1)} \quad 
\frac{1}{2}d^{(3)}_{1,g_1g_2}=\frac{1}{2}d^{(2)}_{2,g_1^2}\quad
\frac{1}{2}d^{(2)}_{1,g_2} = \frac{1}{4} d^{(1)}_{2,g_1} \; ,
\eea
whereas the second duality (\ref{eq:dual2}) implies
\bea
&& d^{(3)}_{q,g_1g_2}= (q+2) \; d^{(2)}_{q,g_2} \quad 
d^{(3)}_{q,g_1^3}=\frac{q+2}{3} \; d^{(2)}_{q,g_1^2} \nonumber\\
&&d^{(2)}_{q,g_1^2}=\frac{q+1}{2} \; d^{(1)}_{q,g_1} \quad
d^{(1)}_{q,g_1}=q \; d^{(0)}_q \; .
\eea

\section{Conclusions}\label{sec:conc}

In this paper we introduced a tower of diagrammatic equations for the planar sector of a square non hermitian random matrix model. We solved the equations in perturbations and proved two duality properties of the solution. 

Our results can be generalized automatically to rectangular non hermitian random matrix models. The generalizations to hermitian matrix models is a bit more subtle as the latter generate also vertices of odd coordination. We are however confindent that one can find corresponding diagrammatic equations related to the quadratic equation obeyed by the resolvent.

Our equation encodes the census of oriented planar diagrams in a very transparent way. Although much more involved, a generalization to graphs of arbitrary genus should be possible.

Another direction of future research is to generalize these equations to group field theories. 
In order to do this several major points still require clarification. First, although we have a proposition for a generalization of the notion of planarity (type 1 graphs in \cite{GFTplanar}), we still need a proof that {\it only} these graphs dominate the partition function. Second, in higher dimmensions, one deals with random tensors, not with random matrices. In order to perform an indepth saddle point approximation one would ideally need to find an appropriate generalization of the notion of eigenvalues for tensors.

\section*{Acknowledgements}

The author would like to thank Jan Ambjorn for suggesting references \cite{Am1,Am2} and for very stimulating discussions regarding the loop insertion equation. Also, the author would like to thank the theoretical physics department at McGill University for hosting him in a impromptu visit at an early stage of this work.

Research at Perimeter Institute is supported by the Government of Cana-da through Industry 
Canada and by the Province of Ontario through the Ministry of Research and Innovation.

\appendix

\section{A simple sum}\label{app:1}

\begin{lemma} Let $S_{q}=\sum_{k=0}^{q} \frac{1}{4^{k+1}} d^{(0)}_k $. Then 
\bea\label{eq:sq}
 S_{q}= \frac{1}{2} \Big{(}1- \frac{1}{4^{q+1}} C^{2q+2}_{q+1} \Big{)}
\eea
\end{lemma}

\noindent{\bf Proof:} Recall that $d^{(0)}_q=\frac{1}{q+1}C^{2q}_q$. We have the Taylor expansion
\bea
 \frac{1}{2}\Big{(}1-\sqrt{1-x} \Big{)}=
\sum_{q=0}^{\infty} \frac{1}{4^{q+1}}d^{(0)}_q x^{q+1} \; ,
\eea
and therefore the generating function of $S_{q}$ is
\bea
 S(x)&=&\sum_{q=0}^{\infty} S_q x^{q+1}= \sum_{q=0}^{\infty} \sum_{k=0}^{q}  \frac{1}{4^{k+1}} 
d^{(0)}_k x^{q+1}=\sum_{k=0}^{\infty} \sum_{q=k}^{\infty}  \frac{1}{4^{k+1}} d^{(0)}_k x^{q+1}\nonumber\\&=&\sum_{k=0}^{\infty} \frac{1}{4^{k+1}} d^{(0)}_k \frac{x^{k+1}}{1-x} 
= \frac{1}{1-x} \sum_{k=0}^{\infty} \frac{1}{4^{k+1}} d^{(0)}_k x^{k+1}\nonumber\\
&=&\frac{1}{2} \Big{(}\frac{1}{1-x}-\frac{1}{\sqrt{1-x}} \Big{)} \; .
\eea
But 
\bea
 \frac{1}{\sqrt{1-x}}=1+\sum_{q=0}^{\infty} \frac{1}{4^{q+1}} C^{2q+2}_{q+1} x^{q+1} \; ,
\eea
hence (\ref{eq:sq}) holds.
\qed

\section{Perturbative computations}\label{app:2}

\subsection{First order}

At first order, eq. (\ref{eq:omzer}), (\ref{eq:os}) and (\ref{eq:T1}) yield
\bea
 \omega^{(1)}(z) = g_1 \Big{(} \frac{1}{2\sqrt{1-\frac{4}{z}}} -\frac{1}{2} 
- \frac{1}{z\sqrt{1-\frac{4}{z}}}\Big{)} \; , 
\eea
and using 
\bea
 \frac{1}{\sqrt{1-\frac{4}{z}}} = \sum_{q=0}^{\infty} \frac{1}{z^q}C^{2q}_q \; ,
\eea
we have
\bea
 \omega^{(1)}(z) &=& g_1 \Big{(}\sum_{q=0}^{\infty} \frac{1}{z^q} \frac{1}{2} C^{2q}_q - \frac{1}{2}
- \sum_{q=0}^{\infty} \frac{1}{z^{q+1}} \frac{1}{2} C^{2q}_q \Big{)} \nonumber\\
    &=& g_1 \sum_{q=0}^{\infty} \frac{1}{z^{q+1}}  
\Big{(}\frac{1}{2}C^{2q+2}_{q+1}-C^{2q}_q \Big{)}\; .
\eea
Therefore
\bea
 d^{(1)}_{q}= g_1 \Big{(}\frac{1}{2}\frac{(2q+2)(2q+1)}{(q+1)^2}-1\Big{)} C^{2q}_q= 
g_1 \frac{q}{q+1} C^{2q}_q \; .
\eea

\subsection{Second order}

At second order we have
\bea
 T^{(2)}(z) &=& g_1\Big{(}\omega^{(1)}(z) - \frac{d^{(1)}_0}{z} \Big{)} +g_2 z
\Big{(}\omega^{(0)}(z)-\frac{d^{(0)}_0}{z} - \frac{d^{(0)}_{1}}{z^2}\Big{)} \nonumber\\
&+&\omega^{(1)}(z) \omega^{(1)}(z)\; ,
\eea
which simplifies to
\bea
 T^{(2)}(z) = g_1 \omega^{(1)}(z) +g_2 z \Big{(}\omega^{(0)}(z)- \frac{1}{z} - \frac{1}{z^2} \Big{)}
+\omega^{(1)}(z) \omega^{(1)}(z) \; .
\eea
We group together the terms containing $g_1$
\bea
 g_1\omega^{(1)}(z)+[\omega^{(1)}(z)]^2 \; ,
\eea
which compute to
\bea
&& g_1^2 \Big{(} \frac{1}{2\sqrt{1-\frac{4}{z}}} 
- \frac{1}{z\sqrt{1-\frac{4}{z}}} -\frac{1}{2} \Big{)} 
\Big{(} \frac{1}{2\sqrt{1-\frac{4}{z}}} 
- \frac{1}{z\sqrt{1-\frac{4}{z}}}  + \frac{1}{2} \Big{)} \nonumber\\
&&= g_1^2 \Big{[}\frac{1}{1-\frac{4}{z}}\Big{(}\frac{1}{2} -\frac{1}{z}\Big{)}^2 -\frac{1}{4}\Big{]}
=g_1^2 \frac{1}{z^2\Big{(}1-\frac{4}{z}\Big{)}} \;,
\eea
thus
\bea
 \omega^{(2)}(z) &=& g_1^2 \frac{1}{z^2\Big{(}1-\frac{4}{z}\Big{)}^{3/2}} \nonumber\\
   &+& g_2 z \Big{(} \frac{1}{2\sqrt{1-\frac{4}{z}}} -\frac{1}{2} 
- \frac{1}{z\sqrt{1-\frac{4}{z}}} -  \frac{1}{z^2 \sqrt{1-\frac{4}{z}}}\Big{)} \; .
\eea
Using 
\bea
 \frac{1}{\Big{(}1-\frac{4}{z}\Big{)}^{3/2}}= \sum_{q=0}^{\infty} \frac{q+1}{2} C^{2q+2}_{q+1} 
\frac{1}{z^{q}} = \sum_{t=1}^{\infty} \frac{t}{2} C^{2t}_{t} 
\frac{1}{z^{t-1}} \; ,
\eea
$\omega^{(2)}(z)$ writes 
\bea
&& g_1^2 \sum_{q=1}^{\infty} \frac{q}{2}C^{2q}_q \frac{1}{z^{q+1}} \nonumber\\
&&+g_2 \Big{(}\sum_{q=0}^{\infty} \frac{1}{2} C^{2q}_q \frac{1}{z^{q-1}} - \frac{z}{2}
- \sum_{q=0}^{\infty} C^{2q}_q \frac{1}{z^{q}}- \sum_{q=0}^{\infty} C^{2q}_q \frac{1}{z^{q+1}}
\Big{)} \; .
\eea
A direct computation shows that the terms $z,z^{0}, z^{-1}$ cancel. Shifting each sum by the appropriate amount yields
\bea
&& g_1^2 \sum_{q=0}^{\infty} \frac{q}{2}C^{2q}_q \frac{1}{z^{q+1}} \nonumber\\
&&+g_2 \Big{(}\sum_{q=0}^{\infty} \frac{1}{2} C^{2q+4}_{q+2} \frac{1}{z^{q+1}} 
- \sum_{q=0}^{\infty} C^{2q+2}_{q+1} \frac{1}{z^{q+1}}- \sum_{q=0}^{\infty} C^{2q}_q \frac{1}{z^{q+1}}
\Big{)}
\;,
\eea
and we conclude
\bea
 d^{(2)}_{q}&=&g_2 \Big{(}\frac{1}{2} C^{2q+4}_{q+2} - C^{2q+2}_{q+1} -  C^{2q}_q \Big{)} 
+g_1^2\frac{q}{2} C^{2q}_{q} \nonumber\\
&=& g_1^2\frac{q}{2} C^{2q}_{q} +g_2 \frac{3q}{q+2} C^{2q}_{q} \; .
\eea

\subsection{Third order}

To the third order we have
\bea\label{eq:ttt}
 T^{(3)}(z) &=& g_1 \omega^{(2)}(z) +g_2  z \Big{(} \omega^{(1)}(z)- \frac{g_1}{z^2}\Big{)}
+g_3 z^2 \Big{(}\omega^{(0)}(z)-\frac{1}{z}-\frac{1}{z^2} -\frac{2}{z^3} \Big{)} \nonumber\\
&+&2\omega^{(1)} \omega^{(2)} \; .
\eea

We start by computing
\bea
 g_1+2\omega^{(1)}=g_1\frac{1}{\Big{(}1-\frac{4}{z}\Big{)}^{1/2}} \Big{(}1-\frac{2}{z} \Big{)} \; ,
\eea
therefore, factoring $\omega^{(2)}$ in the first and last term of (\ref{eq:ttt}) yields
\bea
 T^{(3)}(z)&=&g_1 g_2 \Big{(} \frac{z}{2\sqrt{1-\frac{4}{z}}} -\frac{z}{2} 
- \frac{1}{\sqrt{1-\frac{4}{z}}} -\frac{1}{z} \Big{)} + \nonumber\\
&& +g_3\Big{(} \frac{z^2}{2}-\frac{z^2}{2}\sqrt{1-\frac{4}{z}} - z - 1 - \frac{2}{z}\Big{)} 
\nonumber\\
&&+g_1\frac{1}{\Big{(}1-\frac{4}{z}\Big{)}^{1/2}} \Big{(}1-\frac{2}{z} \Big{)}
\Big{[}g_1^2 \frac{1}{z^2\Big{(}1-\frac{4}{z}\Big{)}^{3/2}} \nonumber\\
&&+ g_2 z \Big{(} \frac{1}{2\sqrt{1-\frac{4}{z}}} -\frac{1}{2} 
- \frac{1}{z\sqrt{1-\frac{4}{z}}} -  \frac{1}{z^2 \sqrt{1-\frac{4}{z}}}\Big{)}\Big{]} 
\; .
\eea

The coefficient of $g_1g_2$ is
\bea
&&\frac{z}{2\sqrt{1-\frac{4}{z}}} -\frac{z}{2} 
- \frac{1}{\sqrt{1-\frac{4}{z}}} -\frac{1}{z} 
+\frac{1}{\Big{(}1-\frac{4}{z}\Big{)}^{1/2}} \Big{(}1-\frac{2}{z} \Big{)} \nonumber\\
&&\times  \Big{(} \frac{z}{2\sqrt{1-\frac{4}{z}}} -\frac{z}{2} 
- \frac{1}{\sqrt{1-\frac{4}{z}}} -  \frac{1}{z \sqrt{1-\frac{4}{z}}}\Big{)} \; ,
\eea
which, after a tedious but straightforeward computation, simplifies to
\bea
\frac{6}{z^2} \frac{1}{1-\frac{4}{z}}
 \; .
\eea

We can now write $\omega^{(3)}$ as
\bea\label{eq:om3}
\omega^{(3)} &=& g_1^{3} \frac{1}{z^2\Big{(}1-\frac{4}{z}\Big{)}^{5/2}} \Big{(}1-\frac{2}{z} \Big{)} \nonumber\\
&+& g_1g_2 \frac{6}{z^2} \frac{1}{\Big{(}1-\frac{4}{z}\Big{)}^{3/2}}\nonumber\\
&+& g_3 \frac{z^2}{\sqrt{1-\frac{4}{z}}} \Big{(}\frac{1}{2}-\frac{1}{2}\sqrt{1-\frac{4}{z}}-\frac{1}{z} -\frac{1}{z^2}-\frac{2}{z^3}\Big{)} 
\; .
\eea

Using 
\bea
 \frac{1}{\Big{(}1-\frac{4}{z}\Big{)}^{5/2}}  =\sum_{t=0}^{\infty} \frac{(t+2)(t+1)}{12}
C^{2t+4}_{t+2} \frac{1}{z^t} \; ,
\eea
the coefficient of $g_1^3$ in (\ref{eq:om3}) is
\bea
&&\sum_{t=0}^{\infty} \frac{(t+2)(t+1)}{12}
C^{2t+4}_{t+2} \frac{1}{z^{t+2}} - \sum_{t=0}^{\infty} \frac{(t+2)(t+1)}{6}
C^{2t+4}_{t+2} \frac{1}{z^{t+3}} \\
&&= \frac{1}{6} C^4_2 \frac{1}{z^2} + \sum_{q=2}^{\infty} \Big{(} \frac{(q+1)q}{12}
C^{2q+2}_{q+1}-  \frac{q(q-1)}{6} C^{2q}_{q} \Big{)} \frac{1}{z^{q+1}} \nonumber\\
&&= \sum_{q=0}^{\infty}\frac{q(q+2)}{6}C^{2q}_q \frac{1}{z^{q+1}}\; , \nonumber
\eea
while the coefficient of $g_1g_2$ in (\ref{eq:om3}) computes
\bea
\frac{6}{z^2} \frac{1}{\Big{(}1-\frac{4}{z} \Big{)}^{3/2} } = 
\sum_{q=0}^{\infty} 6 \frac{q}{2} C^{2q}_{q} \frac{1}{z^{q+1}} \; .
\eea
Finally the coefficient of $g_3$ in (\ref{eq:om3}), 
\bea
\frac{z^2}{2\sqrt{1-\frac{4}{z}}}- \frac{z^2}{2} - z \frac{1}{\sqrt{1-\frac{4}{z}}}-
\frac{1}{\sqrt{1-\frac{4}{z}}}-\frac{2}{z} \frac{1}{\sqrt{1-\frac{4}{z}}}\; ,
\eea
has the Taylor development
\bea \label{eq:om3g3}
- \frac{z^2}{2} + 
 \sum_{q=0}^{\infty} \Big{(}\frac{1}{2} C^{2q}_q \frac{1}{z^{q-2}} - C^{2q}_q \frac{1}{z^{q-1}} - 
C^{2q}_q \frac{1}{z^{q}} - 2 C^{2q}_q \frac{1}{z^{q+1}} \Big{)}\; .
\eea
Again the terms $z^2,z,z^0$ and $z^{-1}$ cancel and shifting each sum by the appropriate amount, eq. (\ref{eq:om3g3}) writes  
\bea
 \sum_{q=0}^{\infty} \Big{(}\frac{1}{2} C^{2q+6}_{q+3}- C^{2q+4}_{q+2} - C^{2q+2}_{q+1} -  2 C^{2q}_q 
\Big{)} \frac{1}{z^{q+1}} = \sum_{q=0}^{\infty}\frac{10 q}{q+3}C^{2q}_q \frac{1}{z^{q+1}} \; .
\eea

Thus, the third order in perturbations is
\bea
 d^{(3)}_{q}=g_1^3 \frac{q(q+2)}{6}C^{2q}_q + g_1g_2 \; 3 q C^{2q}_{q} + g_3 \frac{10 q}{q+3}C^{2q}_q 
\; .
\eea


\begin{thebibliography}{99}

\bibitem{Guhr:1997ve}
  T.~Guhr, A.~Muller-Groeling and H.~A.~Weidenmuller,
  ``Random matrix theories in quantum physics: Common concepts,''
  Phys.\ Rept.\  {\bf 299}, 189 (1998)
  [arXiv:cond-mat/9707301].
  %%CITATION = PRPLC,299,189;%%

\bibitem{Di Francesco:1993nw}
  P.~Di Francesco, P.~H.~Ginsparg and J.~Zinn-Justin,
  ``2-D Gravity and random matrices,''
  Phys.\ Rept.\  {\bf 254}, 1 (1995)
  [arXiv:hep-th/9306153].
  %%CITATION = PRPLC,254,1;%%

\bibitem{Brezin:1977sv}
  E.~Brezin, C.~Itzykson, G.~Parisi and J.~B.~Zuber,
  ``Planar Diagrams,''
  Commun.\ Math.\ Phys.\  {\bf 59}, 35 (1978).
  %%CITATION = CMPHA,59,35;%%

\bibitem{Bouttier:2002iw}
  J.~Bouttier, P.~Di Francesco and E.~Guitter,
  ``Census of Planar Maps: From the One-Matrix Model Solution to a
  Combinatorial Proof,''
  Nucl.\ Phys.\  B {\bf 645}, 477 (2002)
  [arXiv:cond-mat/0207682].
  %%CITATION = NUPHA,B645,477;%%

\bibitem{Di Francesco:2002ru}
  P.~Di Francesco,
  ``Rectangular Matrix Models and Combinatorics of Colored Graphs,''
  Nucl.\ Phys.\  B {\bf 648}, 461 (2003)
  [arXiv:cond-mat/0208037].
  %%CITATION = NUPHA,B648,461;%%

\bibitem{ZinnJustin:2003kw}
  P.~Zinn-Justin and J.~B.~Zuber,
  ``Matrix Integrals and the Generation and Counting of Virtual Tangles and
  Links,''
  J.\ Knot Theor.\ Ramifications {\bf 13}, 325 (2004)
  [arXiv:math-ph/0303049].
  %%CITATION = 00108,13,325;%%

\bibitem{Brezin:1992yc}
  E.~Brezin and J.~Zinn-Justin,
  ``Renormalization group approach to matrix models,''
  Phys.\ Lett.\  B {\bf 288}, 54 (1992)
  [arXiv:hep-th/9206035].
  %%CITATION = PHLTA,B288,54;%%

\bibitem{GrWu1} 
  H.~Grosse and R.~Wulkenhaar,
  ``Renormalisation of phi**4 theory on noncommutative R**4 in the matrix
  base,''
  Commun.\ Math.\ Phys.\  {\bf 256}, 305 (2005)
  [arXiv:hep-th/0401128].
  %%CITATION = CMPHA,256,305;%%

\bibitem{GrWu2}
  H.~Grosse and R.~Wulkenhaar,
  ``Power-counting theorem for non-local matrix models and renormalisation,''
  Commun.\ Math.\ Phys.\  {\bf 254}, 91 (2005)
  [arXiv:hep-th/0305066].
  %%CITATION = CMPHA,254,91;%%

\bibitem{DGMR}
  M.~Disertori, R.~Gurau, J.~Magnen and V.~Rivasseau,
  {\it ``Vanishing of beta function of non commutative phi(4)**4 theory to all
  orders,''}
  Phys.\ Lett.\  B {\bf 649}, (2007) 95
  [arXiv:hep-th/0612251].
  %%CITATION = PHLTA,B649,95;%%

\bibitem{GGR}
  J.~B.~Geloun, R.~Gurau and V.~Rivasseau,
  {\it ``Vanishing beta function for Grosse-Wulkenhaar model in a magnetic field,''}
  Phys.\ Lett.\  B {\bf 671}, 284 (2009)
  [arXiv:0805.4362 [hep-th]].
  %%CITATION = PHLTA,B671,284;%%

\bibitem{Am1}
   J.~Ambjorn, C.~F.~Kristjansen and Yu.~M.~Makeenko,
   ``Higher Genus Correlators For The Complex Matrix Model,''
   Mod.\ Phys.\ Lett.\  A {\bf 7} (1992) 3187
   [arXiv:hep-th/9207020].
  %%CITATION = MPLAE,A7,3187;%%

\bibitem{Am2}
   J.~Ambjorn, L.~Chekhov, C.~F.~Kristjansen and Yu.~Makeenko,
   ``Matrix model calculations beyond the spherical limit,''
   Nucl.\ Phys.\  B {\bf 404} (1993) 127
   [Erratum-ibid.\  B {\bf 449} (1995) 681]
   [arXiv:hep-th/9302014].
   %%CITATION = NUPHA,B404,127;%%

\bibitem{laurentgft}
  L.~Freidel,
  ``Group field theory: An overview,''
  Int.\ J.\ Theor.\ Phys.\  {\bf 44}, 1769 (2005)
  [arXiv:hep-th/0505016].
  %%CITATION = IJTPB,44,1769;%%


\bibitem{iogft} D. Oriti, in {\it Quantum Gravity}, B. Fauser, J. Tolksdorf and E. Zeidler, eds., Birkhaeuser, Basel, (2007), [arXiv: gr-qc/0512103]

\bibitem{Color}
  R.~Gurau,
  {\it ``Colored Group Field Theory''},
  [arXiv:0907.2582 [hep-th]].
  %%CITATION = ARXIV:0907.2582;%% 

\bibitem{PolyColor}
   R.~Gurau,
  ``Topological Graph Polynomials in Colored Group Field Theory,''
  arXiv:0911.1945 [hep-th].
  %%CITATION = ARXIV:0911.1945;%%

\bibitem{GFTplanar}
  L.~Freidel, R.~Gurau and D.~Oriti,
  ``Group field theory renormalization - the 3d case: power counting of
  divergences,''
  Phys.\ Rev.\  D {\bf 80}, 044007 (2009)
  [arXiv:0905.3772 [hep-th]].
  %%CITATION = PHRVA,D80,044007;%%

\bibitem{Sefu1}
  J.~Magnen, K.~Noui, V.~Rivasseau and M.~Smerlak,
  ``Scaling behaviour of three-dimensional group field theory,''
  Class.\ Quant.\ Grav.\  {\bf 26}, 185012 (2009)
  [arXiv:0906.5477 [hep-th]].
  %%CITATION = CQGRD,26,185012;%%

\bibitem{Sefu2}
  J.~B.~Geloun, T.~Krajewski, J.~Magnen and V.~Rivasseau,
  ``Linearized Group Field Theory and Power Counting Theorems,''
  arXiv:1002.3592 [hep-th].
  %%CITATION = ARXIV:1002.3592;%%

\bibitem{Samhof} M. Salmhofer,  Renormalization: An Introduction, Springer; ISBN-10: 3540646663, ISBN-13: 978-3540646662

\bibitem{Polchinski:1983gv}
  J.~Polchinski,
  ``Renormalization And Effective Lagrangians,''
  Nucl.\ Phys.\  B {\bf 231}, 269 (1984).
  %%CITATION = NUPHA,B231,269;%%

\bibitem{Morris:1990cq}
  T.~R.~Morris,
  ``Checkered surfaces and complex matrices,''
  Nucl.\ Phys.\  B {\bf 356}, 703 (1991).
  %%CITATION = NUPHA,B356,703;%%



































\end{thebibliography}
\end{document}